# Magnetic properties of icosahedral quasicrystals and their cubic approximants in the Cd-Mg-RE (RE = Gd, Tb, Dy, Ho, Er, and Tm) systems

Farid Labib[1], Daisuke Okuyama[1], Nobuhisa Fujita[1], Tsunetomo Yamada[2], Satoshi Ohhashi[1], Taku J. Sato[1] and An-Pang Tsai[1]

[1] Institute of Multidisciplinary Research for Advanced Materials (IMRAM), Tohoku University, Sendai 980-8577, Japan
[2] Faculty of Science, Department of Applied Physics, Tokyo University of Science, Katsushika-ku, Tokyo, Japan

E-mail: labib.farid.t2@dc.tohoku.ac.jp

## Abstract

A systematic investigation has been performed to elucidate effects of Rare-Earth (RE) type and local atomic configuration on magnetic properties of icosahedral quasicrystal (iQC) and their cubic approximants (2/1 and 1/1 ACs) in the ternary Cd-Mg-RE (RE = Gd, Tb, Dy, Ho, Er, and Tm) systems. At low temperatures, iQC and 2/1 ACs exhibit spin-glass-like freezing for RE = Gd, Tb, Dy, and Ho, while Er and Tm systems do not show freezing behaviour down to the base temperature ~ 2 K. The 1/1 ACs exhibit either spin-glass-like freezing or antiferromagnetic (AFM) ordering depending on their constituent Mg content. The $T_f$ values show increasing trend from iQC to 2/1 and 1/1 ACs. In contrast, the absolute values of Weiss temperature for iQCs are larger than those in 2/1 and 1/1 ACs, indicating that the total AFM interactions between the neighboring spins are larger in aperiodic, rather than periodic systems. Competing spin interactions originating from the long-range Ruderman-Kittel-Kasuya-Yoshida mechanism along with chemical disorder of Cd/Mg ions presumably account for the observed spin-glass-like behavior in Cd-Mg-RE iQCs and ACs.

## 1. Introduction

Icosahedral quasicrystals (iQC), as aperiodically-ordered intermetallic compounds, generate sharp Bragg reflections with 5-fold rotational symmetry, indicating the presence of a long-range positional order without periodicity in their atomic configuration [1,2]. Based on their atomic structure, iQCs can be classified into three subclasses, namely the Mackay- [3], Bergman- [4] and Tsai-type [5]. The Tsai-type subclass, in particular, comprising the binary i-CdREs (rare-earth) [6], ternary i-CdMgREs [7,8], i-AgInREs (RE = Yb) [9], i-AuAlREs (RE = Yb, Tm) [10,11], and so forth might be considered as the largest of the three. Figure1a illustrates shell structure of the Tsai-type iQCs [12,13]. Magnetic properties of the Tsai-type iQCs in the Cd-RE and Cd-Mg-RE systems with trivalent REs (i.e. excluding Yb) have been investigated in a number of studies [14–19]. The results are summarized as follows: the temperature dependence of the magnetization follow the Curie-Weiss law at high-temperatures. The effective magnetic moments are consistent with free $RE^{3+}$ ions. At low temperatures, iQCs in Cd-RE and Cd-Mg-RE exhibit spin-glass-like freezing behavior. The short-range-magnetic correlations in the spin-glass-like state has been observed in the $A$-Mg-RE ($A$ = Zn or Cd) QCs at low temperatures using neutron scattering [14]. Magnetic properties of the iQCs (in both binary Cd-RE and ternary Cd-Mg-RE systems) are summarized in Table 1.

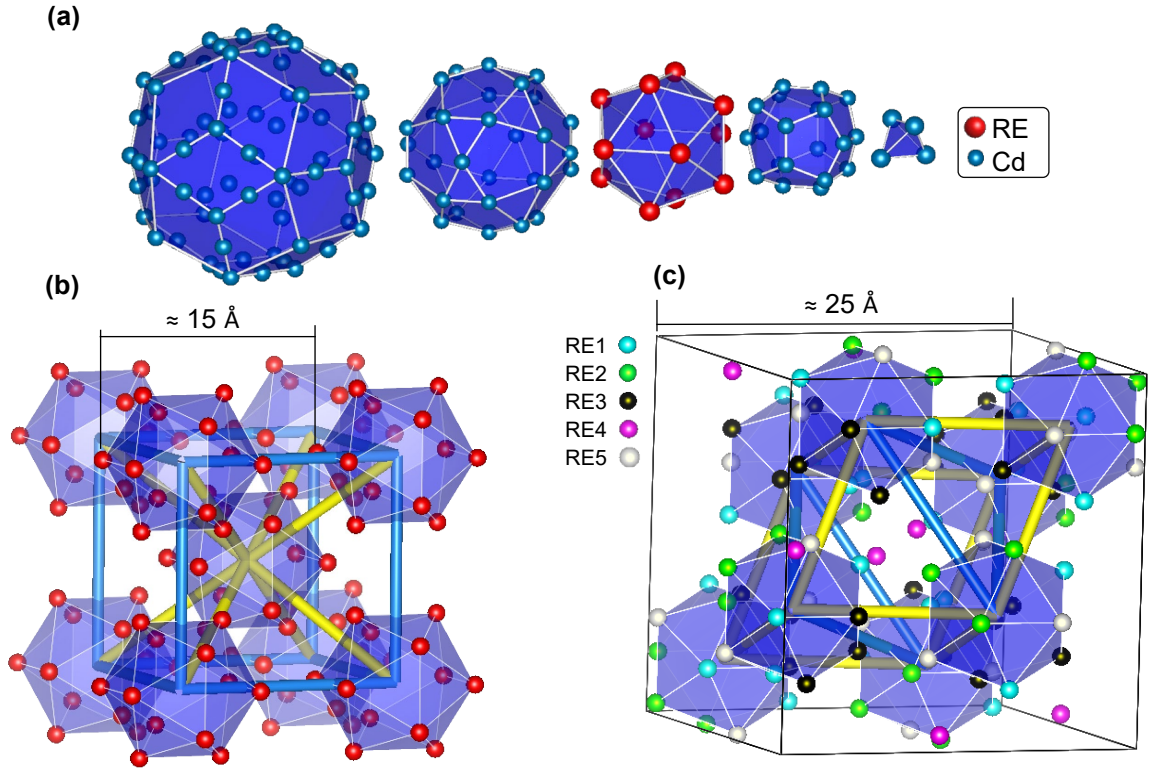

**Figure 1.** (a) Shell structure of the Tsai-type icosahedral quasicrystal (iQC) in the binary Cd-Tb system. From left to right: Cd rhombic triacontahedron (RTH) (92 atoms), Cd icosidodecahedron (30 atoms), RE icosahedron (12 atoms), Cd dodecahedron (20 atoms), and inner Cd tetrahedron (4 atoms). (b,c) RE sites on the icosahedron shell of the RTH clusters in the atomic structures of 1/1 and 2/1 ACs in Cd-Mg-RE systems. 2- and 3-fold linkages between the cluster centres are represented by thick light blue and yellow bonds, respectively, in Figures 1b and 1c.

In a close proximity of iQCs yet with slightly different compositions, the same rhombic tricontahedron (RTH) clusters can be arranged with translational symmetry generating their crystalline counterpart, so-called approximants crystals (ACs) [13]. With compared to the 1/1 AC (with a space group of $Im\bar{3}$), the structure of the 2/1 AC ($Pa\bar{3}$) is closer to the iQC ($Pm\bar{3}5$) [20,21]. The configurations of RE atoms within the unit cells of the 1/1 and 2/1 ACs are illustrated schematically in Figures 1b and 1c, respectively. Note that in the former, there exist 24 RE sites that are symmetrically equivalent. They form the icosahedral shells of the RTH clusters. In the latter, on the other hand, the 104 RE sites are divided into five Wyckoff orbits, represented by spheres of different colours in Figure 1c, all of which except RE4 belong to the icosahedral shells of the RTH clusters. The four dimers of RE4 sites are arranged on the long body diagonals of the four acute rhombohedron units which fill in the gaps between the RTH clusters [12].

As represented in Figure 2, the shortest RE-RE distances in both 1/1 and 2/1 ACs (within a range of 3.4 to 6.2Å) can be classified into the following seven categories: (I) icosahedral edge, (II) side edge of trigonal antiprisms, each of which has base triangles belonging to an adjacent pair of icosahedra aligned along a 3-fold axis, (III) side edge of rectangles, each of which connects parallel edges from an adjacent pair of icosahedra aligned along a 2-fold axis. For the 2/1 ACs, additional categories are required: (IV-1, IV-2, and IV-3) three distinct distances between RE4 sites to icosahedral sites, and (V) short RE4-RE4 distance for dimers, associated with acute rhombohedron units. The 2- and 3-fold linkages between the cluster centres are represented by thick light blue and yellow bonds, respectively, in Figures 1b and 1c.

Magnetic properties of the Tsai-type 1/1 ACs have been investigated in several systems. While some of the ternary 1/1 approximants show spin-glass-like freezing, the binary $Cd_6RE$ [15,22–24] and ternary Au-$SM$-RE (RE = Gd and $SM$ = Si, Ge) compounds [25–27] are of special interest since their magnetic ordering are not of a spin-glass type but are of a long-range-antiferromagnetic (AFM) and ferromagnetic (FM) ordering, respectively. The binary Cd$_6$RE compounds, in particular, exhibit one or several anomalies in their magnetic susceptibilities, see table 1, indicating the existence of several distinct AFM phases [22,23].

**Table 1.** Magnetic properties of the binary Cd-RE and ternary Cd-Mg-RE (RE: rare earth) icosahedral quasicrystals (iQCs) and approximant crystals (ACs). $\Theta_w$ is Weiss temperature, $T_f$ is the spin freezing temperature, $\mu_{eff}$ is the experimental effective moment and $\mu_{calc}$ is the calculated free-ion value

| iQC/AC | System | $\mu_{eff}$ ($\mu_B/RE_{ion}$) | $\mu_{calc.}$ ($\mu_B/RE_{ion}$) | $\Theta_w$ (K) | $T_f$ (K) | $T_{N1}$ (K) | $T_{N2}$ (K) | $T_{N3}$ (K) | $T_{N4}$ (K) | Ref. |
|---|---|---|---|---|---|---|---|---|---|---|
| i-QC | Cd-Mg-Gd | 7.24 | | -37.8 | 4.3 | – | – | – | – | [34] |
| i-QC | Cd$_{7.88}$Gd | – | 7.94 | -41 | 4.6 | – | – | – | – | [18] |
| 1/1 AC | Cd$_6$Gd | 7.94 | | -32 | – | 18.9 | 13.2 | 7.3 | 2.5 | [23] |
| i-QC | Cd-Mg-Tb | 9.74 | | -24.5 | 5.9 | – | – | – | – | [34,16] |
| i-QC | Cd$_{7.69}$Tb | – | 9.72 | -21 | 5.3 | – | – | – | – | [18] |
| 1/1 AC | Cd$_6$Tb | 9.8 | | -17 | – | 24 | 19 | 2.4 | – | [22,23] |
| i-QC | Cd-Mg-Dy | 10.59 | | -18.4 | 3.2 | – | – | – | – | [34] |
| i-QC | Cd$_{7.50}$Dy | – | 10.63 | -11 | 3 | – | – | – | – | [18] |
| 1/1 AC | Cd$_6$Dy | 10.9 | | -5.1 | – | 17.8 | – | – | – | [23] |
| i-QC | Cd-Mg-Ho | 10.42 | | -7 | 12.5 | – | – | – | – | [34] |
| i-QC | Cd$_{7.60}$Ho | – | 10.60 | -6 | 1.76 | – | – | – | – | [18] |
| 1/1 AC | Cd$_6$Ho | 10.5 | | -1 | – | 8.4 | 6.8 | 3.4 | – | [23] |
| i-QC | Cd-Mg-Er | 9.71 | | -6 | 4.4 | – | – | – | – | [34] |
| i-QC | Cd$_{7.34}$Er | – | 9.59 | -4 | 1.11 | – | – | – | – | [18] |
| 1/1 AC | Cd$_6$Er | 9.1 | | -0.9 | – | 2.8 | – | – | – | [23] |
| i-QC | Cd-Mg-Tm | 7.08 | | -2 | – | – | – | – | – | [34] |
| i-QC | Cd$_{7.28}$Tm | – | 7.57 | -2 | 0.63 | – | – | – | – | [18] |
| 1/1 AC | Cd$_6$Tm | 7.4 | | -3.1 | – | 2.2 | – | – | – | [23] |

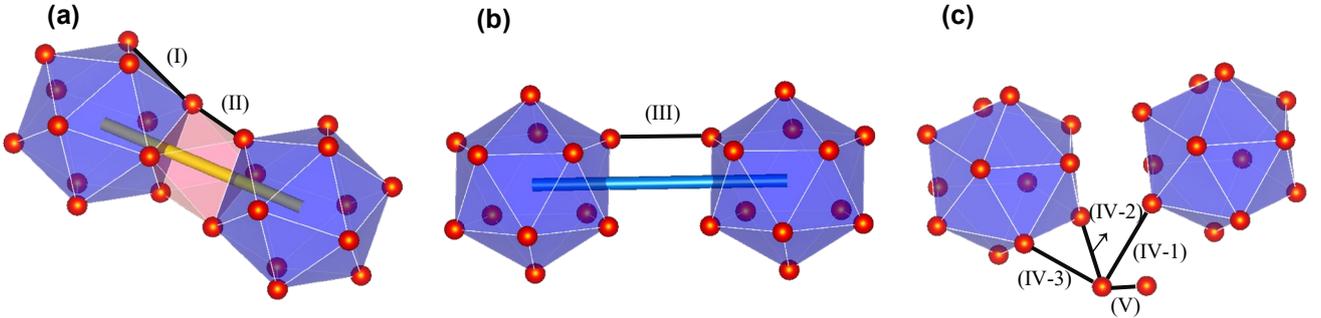

**Figure 2.** Seven categories of shortest RE-RE distances (within a range of 3.4 to 6.2Å). (a) An edge of an icosahedron (I) and a side edge of a trigonal antiprism connecting triangular faces of adjacent icosahedra aligned along a 3-fold axis (II). (b) A side edge of a rectangle connecting parallel edges from adjacent icosahedra aligned along a 2-fold axis (III). (c) Three kinds of distances between a RE4 site to icosahedral sites (IV-1, IV-2, and IV-3) as well as a short distance within a RE4-RE4 dimer (V). (c) is only for the 2/1 AC, whereas (a) and (b) are common. The typical values of distances for the Cd-Mg-Er 2/1 AC are as follows: (I): 5.73 Å, (II): 5.75 Å, (III): 6.12 Å, (IV-1): 6.04 Å, (IV-2): 5.64 Å, (IV-3): 5.59 Å, and (V): 3.47 Å.

To the best of our knowledge, no previous study has investigated magnetic properties of the ACs in the ternary Cd-Mg-RE systems. One of unique features of the ternary Cd-Mg-RE (RE = Gd, Tb, Dy, Ho, Er, and Tm) alloy systems is the opportunity to obtain iQCs and their successive cubic 2/1 and 1/1 ACs in the same set of alloy systems [28,29]. The systematic investigation of the iQCs, 2/1 and 1/1 ACs should make an important contribution to better understand the nature of magnetic freezing in aperiodic systems. The present research, therefore, intends to perform a systematic magnetic susceptibility measurements on iQC, 2/1 and 1/1 ACs in the Cd-Mg-RE (RE = Gd, Tb, Dy, Ho, Er, and Tm) systems to show if there is any trend, which may further serve as a guideline for tuning or tailoring the magnetic properties.

## 2. Experiment

As-cast Cd-Mg-RE (RE = Gd, Tb, Dy, Ho, Er, and Tm) alloys were prepared by encapsulating three grams of starting elements into stainless-steel tubes employing arc-welding under Ar atmosphere. The tubes were further placed inside quartz tubes under depressurized Ar gas (~550 Torr). Based on the developed pseudo-binary phase diagram [28,29], the ACs are stable at ~ 773 K, while the iQCs are stabilized at lower temperatures. Except the 2/1 ACs that have very strict compositions (i.e., $Cd_{65}Mg_{22}RE_{13}$), both the 1/1 ACs and iQCs occur inside relatively large and elongated single-phase domains with varying Cd/Mg ratios in the ternary phase diagrams. Such a high solubility of Mg in the Cd-based systems is usually discussed by almost similar atomic size and valence number of Mg and Cd whereby the replacement of Cd by Mg would thus not cause much strain or change in electron concentration. However, for unknown reasons, the single domains of the 1/1 ACs, i.e., $(Cd,Mg)_6RE$, behave differently depending on the RE elements. The largest domain is observed for RE = Tb, where it spans up to $Cd_{65}Mg_{20}RE_{15}$ in the ternary phase diagram. For RE = Dy, Ho and Er, on the other hand, the single 1/1 AC domains shrink to $\approx$ 10 at.% Mg. Given that the 2/1 AC forms inside a very strict compositional area close to $Cd_{65}Mg_{22}RE_{13}$, we tried to keep the Mg content of the iQCs and 1/1 ACs as close as possible to $\approx$ 20 - 22 at.%. Consequently, alloys with nominal compositions of $Cd_{65}Mg_{23}RE_{12}$, $Cd_{65}Mg_{22}RE_{13}$ and $Cd_{65+x}Mg_{20-x}RE_{15}$ ($x = 0$ for RE = Tb, and x = 10 for RE = Dy, Ho, Er, Tm) were carefully selected to synthesize single-phase polycrystalline iQC, 2/1 and 1/1 ACs, respectively. Despite several attempts, no 2/1 AC but 1/1 AC is detected in an alloy with nominal composition of $Cd_{65}Mg_{22}Gd_{13}$. Investigating the effects of Mg addition on magnetic behaviours of the 1/1 ACs falls outside the scope of the current paper and thus is left for future work. The heat treatment protocols for synthesizing iQC and ACs are as follows: after an initial melting of the prepared alloys at 973 K, iQC and ACs were isothermally annealed at 673 K for over 140 h and 773 K for over 100 h, respectively. Microstructures and local compositions of the prepared samples were analysed by scanning electron microscopy (SEM) equipped with energy dispersive X-ray (EDX) spectrometer. The specimens for TEM observation were prepared using 'crush-and-float' method during which the samples were crushed in ethanol and lightly floated pieces were then transferred to a Cu grid. For electron backscatter diffraction (EBSD) analysis, the surface of the samples was polished by ion bombardment machine with accelerated voltage, gun current, milling time and ion angle of 2 kV, 2 mA, 3 h, and 12°, respectively. Powder X-ray diffractometry (XRD; Mac science M03XHF22) with Cu-$K_\alpha$ was used for the phase identification of the prepared alloys. The temperature dependence of the dc magnetic susceptibility was measured in a temperature range of 2 – 300 K under 100 Oe using superconducting quantum interfering device (SQUID) magnetometer (Quantum Design, MPMS-XL). The data collection was conducted for both zero-field cooled (ZFC) and field cooled (FC) conditions. The ac magnetic susceptibility measurements were carried out for frequencies varying from 0.1 to 10 Hz in the temperature range of 2 – 20 K under zero external magnetic field.

## 3. Results and discussion

Typical backscattering SEM images from microstructures of the $Cd_{65}Mg_{23}Tb_{12}$, $Cd_{65}Mg_{22}Tb_{13}$ and $Cd_{65}Mg_{20}Tb_{15}$, which can be assigned as single-phase iQC, 2/1 and 1/1 ACs, are presented in Figures 3a-c. The SEM analysis of the specified regions are shown on top right corner of the images. As seen, the analysed compositions are quite consistent with the nominal values. The atomic concentration of the magnetic Tb shows gradual increase from 11.97 at% in iQC to 14.35 at% in 1/1 AC. The corresponding EBSD Kikuchi patterns are displayed in Figures 3d-f. As shown, the iQC and 2/1 AC exhibit almost undistinguishable patterns with obvious pentagon-shaped Kikuchi bands, at the center of which the bands intersect at a unique point (marked as 5-fold [211111] and pseudo-5-fold [805] in iQC and 2/1 AC, respectively). This indicates high resemblance in atomic structures of the iQC and 2/1 AC. However, the Kikuchi pattern obtained from the 1/1 AC reveals split Kikuchi bands forming a distorted pentagonal band with pseudo-5-fold [503] pole at the center suggesting a significant deviation of the 1/1 AC structure from the iQC one. Figures 3g-i provide selected area electron diffraction (SAED) patterns of the iQC and ACs taken with incidences along their 3-fold axes. As seen, while the iQC represents sharp diffraction spots inflating in $\tau$ order, the pattern obtained from the 2/1 AC in Figure 3h exhibits zig-zag array of diffraction peaks (indicated by arrows) originating from a linear phason strain in the atomic structure [30]. The SAED pattern of the 1/1 AC in Figure 3i represents a larger displacement of diffraction spots from ideal positions indicating a larger magnitude of linear phason strain in the structure compared to the 2/1 AC. The results provided in Figure 3 evidence a systematic and successive deviation of the atomic structure in, respectively, 2/1 and 1/1 AC from an ideal iQC symmetry. This is of special interest in the present study since it offers a unique opportunity to obtain a possible trend in their magnetic behavior by performing magnetic susceptibility measurements.

Figure 4 depicts the edge length of a surface rhombi in the RTH cluster ($a_q$), definition of which is shown in the inset of figure for clarity, obtained from iQC, 2/1 and 1/1 ACs in Cd-Mg-RE systems. While $a_q$ is the quasilattice constant for iQC, $a_q$ in ACs may be calculated from their measured lattice parameters ($a_{q/p}$) by neglecting possible distortion of the RTH cluster due to slightly different local atomic configuration in iQC and ACs, using the following equation:

$$a_{q/p} = 2\sqrt{\frac{1}{2+\tau}(p+q\tau)} a_q \qquad (1)$$

where $p$ and $q$ indicates AC indices, e.g. $p = 1$, $q = 2$ for 2/1 AC, and $\tau$ is the golden mean [20]. Clearly, $a_q$ increases from 5.588 Å for the CdMgTm 2/1 AC to 5.652 Å for i-CdMgGd.

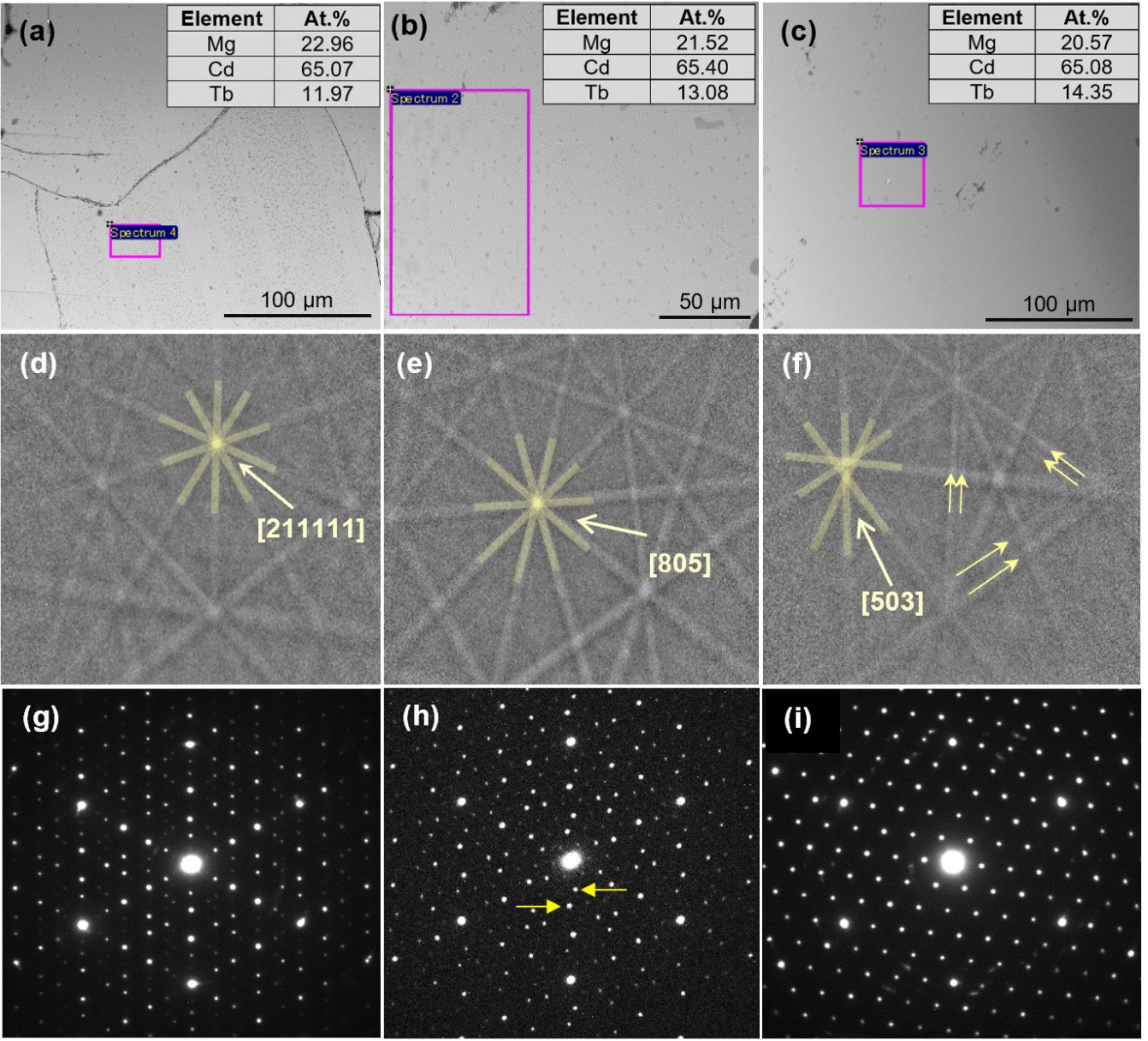

**Figure 3.** (a-c) Backscattered scanning electron microscopy (SEM) images, (d-f) electron backscatter diffraction (EBSD) Kikuchi patterns and (g-i) selected area electron diffraction (SAED) patterns taken with incidences along 3-fold axes of (a, d, g) iQC, (b, e, h) 2/1 AC and (c, f, i) 1/1 AC in the Cd-Mg-Tb system. The energy dispersive X-ay (EDX) analysis of specific regions are shown on top right corner of the figures (a-c).

Provided that the RTH structure is identical, such increment of $a_q$ in i-CdMgREs can be viewed as the effect of RE atomic size on the expansion of the RTH cluster.

Figure 5 presents the temperature dependence of the magnetic susceptibility ($x_{dc}$) and inverse magnetic susceptibility ($1/x_{dc}$) for the iQC and ACs measured under 100 Oe in the temperature range of 2–300 K. The magnified low-temperature susceptibilities from 2 to 30 K are also shown in the inset. Open and close circles in Figure 5 represent FC and ZFC curves, respectively. At high temperatures (100 K < $T$ < 300 K), the temperature dependence of inverse magnetic susceptibility in all samples falls on a line following the Curie-Weiss law:

$$\chi(T) = \frac{N_A \mu_{\text{eff}}^2 \mu_B^2}{3k_B(T-\theta_w)} \quad (2)$$

where $N_A$, $k_B$, $\mu_B$, $\mu_{\text{eff}}$ and $\theta_w$ are Avogadro's number, the Boltzmann factor, Bohr magneton, effective magnetic moment and Weiss temperature, respectively. $\theta_w$ values representing a sum of all exchange interactions are obtained by extrapolating the least-squares fitting of the susceptibility data to the temperature axis. The freezing temperature $T_f$ was obtained from the cusp of the ZFC dc magnetic susceptibility. The transition temperature $T_N$ for AFM ordering was also obtained from the cusp at the ZFC and FC dc magnetic susceptibilities. Table 2 summarizes the estimated $\mu_{eff}$, $\theta_w$, and $T_f$ or $T_N$. The uncertainty in estimation of $\theta_w$ corresponds to standard deviation arising from least-squares fitting of the susceptibility data over different temperature ranges. As seen, the effective moments $\mu_{eff}$ are close to the calculated values for free $RE^{3+}$ ions, $\mu_{RE^{3+}} = g\sqrt{J(J+1)}\mu_B$ ($g$ and $J$ stand for the Landé $g$-factor and total magnetic angular moment, respectively), indicating that RE atoms are trivalent. Moreover, all samples exhibit negative $\theta_w$ values, indicating that RE-RE exchange interactions are dominantly antiferromagnetic.

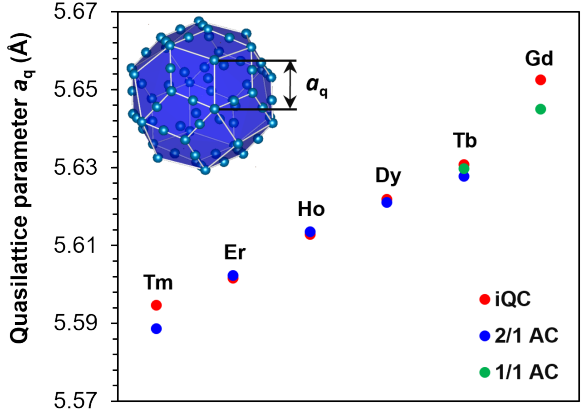

**Figure 4.** Variation of the quasilattice parameter ($a_q$), defined as the edge length of the RTH clusters in Cd-Mg-$RE$ iQCs ($Cd_{65}Mg_{23}RE_{12}$), 2/1ACs ($Cd_{65}Mg_{22}RE_{13}$) and 1/1 ACs ($Cd_{65}Mg_{20}Tb_{15}$ and $Cd_{65}Mg_{22}Gd_{13}$) consisting 20–23 at. % Mg.

At low temperatures, as shown in the insets of Figure 5, the magnetic susceptibility of the iQC and 2/1 ACs for the RE atoms except Tm and Er exhibits bifurcation between the ZFC and FC curves below $T_f$. Such bifurcation is associated with the spin-glass-like freezing. For further confirmation of the spin-freezing phenomenon, the temperature dependence of the ac magnetic susceptibility of the 2/1 AC in the Cd-Mg-Ho system was measured under selected frequencies between 1 to 100 Hz. The result is shown in Figure 6a. The position of the cusp ($T_f$) is shifted about 0.6 K to the higher temperatures, and the magnitude is reduced by 9.43 %, as the frequency is increased from 1 to 100 Hz. The imaginary part $\tilde{\chi}_{ac}$ in Figure 6b illustrates a sharp rise near $T_f$. Such a frequency dependent variation of the $\chi'_{ac}$ and $\tilde{\chi}_{ac}$ offers a reliable criteria for distinguishing a spin-glass-like material from other magnetic systems [31,32]. Note that this is the first observation of the spin-glass-like feature in the cubic 2/1 and 1/1 ACs in the ternary Cd-Mg-RE systems. In the case of iQCs (RE = Ho and Er), 2/1 and 1/1 ACs (RE = Er and Tm) no cusp has been observed down to 2 K.

In Figure 5, some samples such as iQC and 1/1 AC in Cd-Mg-Tb system exhibit relatively broader cusp in both ZFC and FC susceptibility curves compared to other systems. Such behavior accords with earlier observations in Tb-contained binary i-CdREs [18] and ternary i-CdMgREs [16] and may possibly originate from a gradual blocking of the AFM cluster rotations causing the magnetic system to slowly fall into the frozen state. Sebastian et al. [16], however, associated such behavior with a possible presence of magnetic impurities such as oxides. Moreover, magnetic susceptibilities in Figures 5c and 5e evidence separation of the FC and ZFC curves at higher temperatures than the $T_f$ for iQC and ACs in Cd-Mg-RE (RE = Tb and Dy) systems. This behavior is also consistent with earlier reports in binary i-Cd-Dy/Tb [6] and ternary i-Cd-Mg-Tb [16] systems and is attributed to the intrinsic tendency of these compounds to have a distribution in freezing temperature [14]. In Figures 5e and 5g, the low temperature magnetic susceptibility of the 1/1 ACs in Cd-Mg-RE (RE = Dy and Ho) systems is shown. It does not exhibit spin-glass-like bifurcation of FC and ZFC but a conventional AFM ordering; one sharp anomaly at $T_N = 7$ K in $Cd_{75}Mg_{10}Ho_{15}$ and two anomalies at $T_N = 5$ and 11 K in $Cd_{75}Mg_{10}Dy_{15}$, are clearly seen in Figures 5e and 5g. The magnetic ordering in these compounds may originate from their lower Mg content (i.e. 10 at. %), compared to the rest of the alloys that contain 20-23 at% Mg (see Table 2). One may expect substantially less chemical disorder for the Cd/Mg mixed sites, and this less disorder may suppress the spin-glass freezing, resulting in the AFM ordering, as commonly found in the binary (chemically non-disordered) $Cd_6RE$ compounds [22,23]. Such observation suggests chemical-disorder-driven AFM to spin-glass transition that presumably occurs as the Mg content of 1/1 ACs increases. A full discussion of them lies beyond the scope of the current study and will be dealt with in a separate article, which is under preparation [33].

Next, to perform the systematic analysis, we show $T_f$ (or $T_N$ if AFM order is established) and $\theta_w$ as a function of de Gennes factor, $\xi = (g-1)^2 J(J+1)$, depicted in Figures 7a and 7b. The $|\theta_w|$ values for iQC and ACs are approximately proportional to the de Gennes factor of the magnetic RE, which is in agreement with several earlier reports in other magnetic iQCs [18,23,26,34]. On the other hand, the scaling for $T_f$ is only approximate; serious deviation can be found for RE = Gd. This is most likely due to the difference in the single ion anisotropy; because of the vanishing orbital component, $Gd^{3+}$ is exceptionally isotropic among other RE ions in crystalline electric field, giving rise to the lower transition temperature. In addition to the de Gennes scaling, the $T_f$ or $T_N$ values in Figures 7a and 7b show contrasting trends from iQC to 2/1 and further to the 1/1 ACs containing the same type RE element. Exemplified by RE = Tb, $T_f$ is largest for 1/1 AC, while the smallest $T_f$ is found for iQC. In striking contrast, the $|\theta_w|$ values for iQCs are larger than those in 2/1 and 1/1 ACs, respectively suggesting that the total AFM interaction is increasing from 1/1 AC to 2/1 AC and further to the iQC. This indicates that the empirical $|\Theta_w/T_f|$ frustration parameter [34], is the largest for iQC, and shows decreasing behavior as the quasiperiodicity is replaced by its rational approximation. The $|\Theta_w/T_f|$ yields 3 – 8 times larger values for iQCs than ACs (see Table 2). The same conclusion can also be drawn by careful inspection of the magnetic parameters for the binary Cd-RE iQCs and 1/1 ACs, as summarized in Table 1.

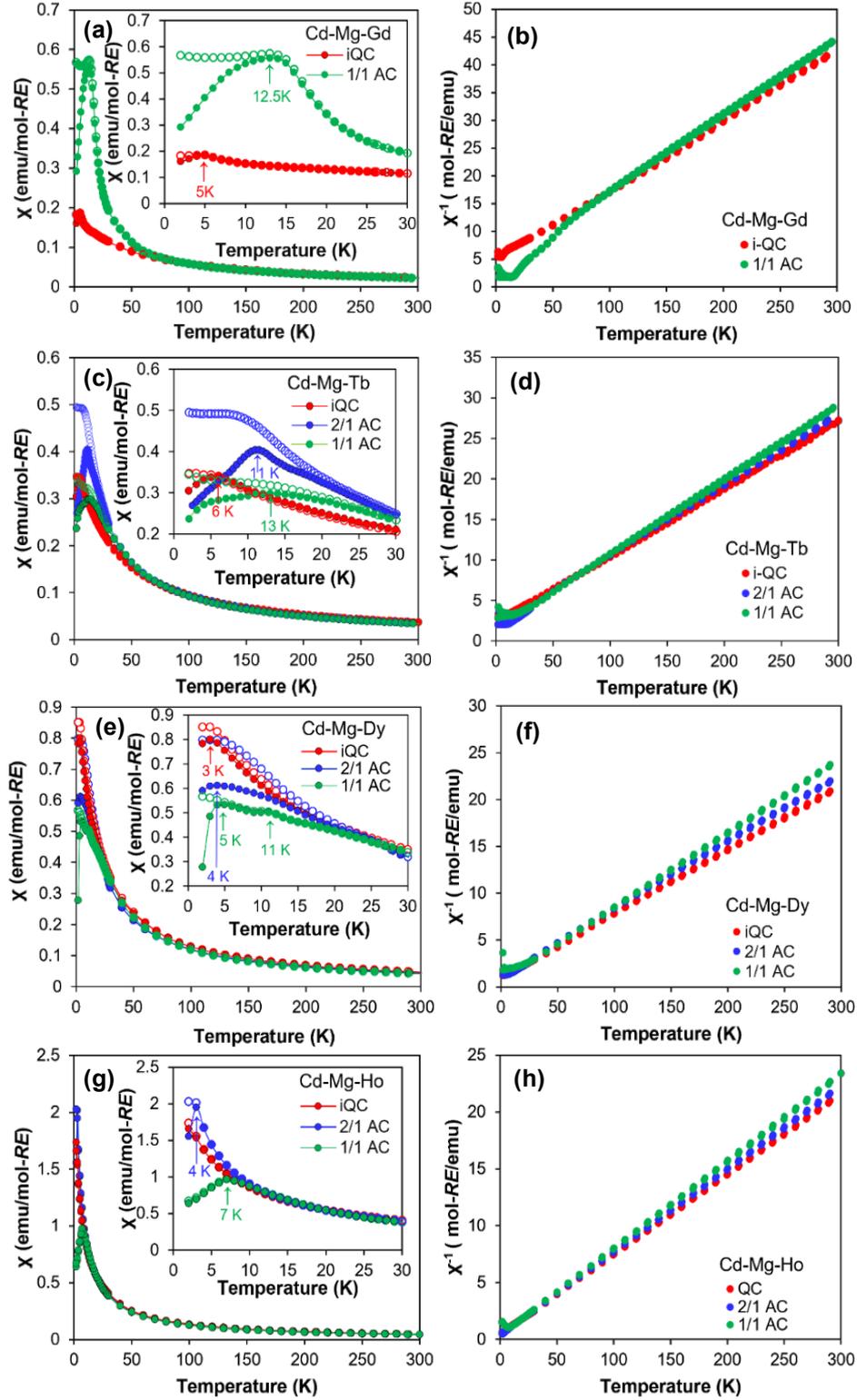

**Figure 5.** Temperature dependence of the (a, c, e, g, j, l) magnetic susceptibilities and (b, d, f, h, k, m) inverse magnetic susceptibilities of the i-QC, 2/1 and 1/1 AC in the Cd-Mg-*RE* (*RE* = Gd, Tb, Dy, Ho, Er and Tm) systems measured under $H$ = 100 Oe showing both field cooled (FC) and zero field cooled (ZFC) behaviours. Open and close circles represent FC and ZFC curves, respectively.

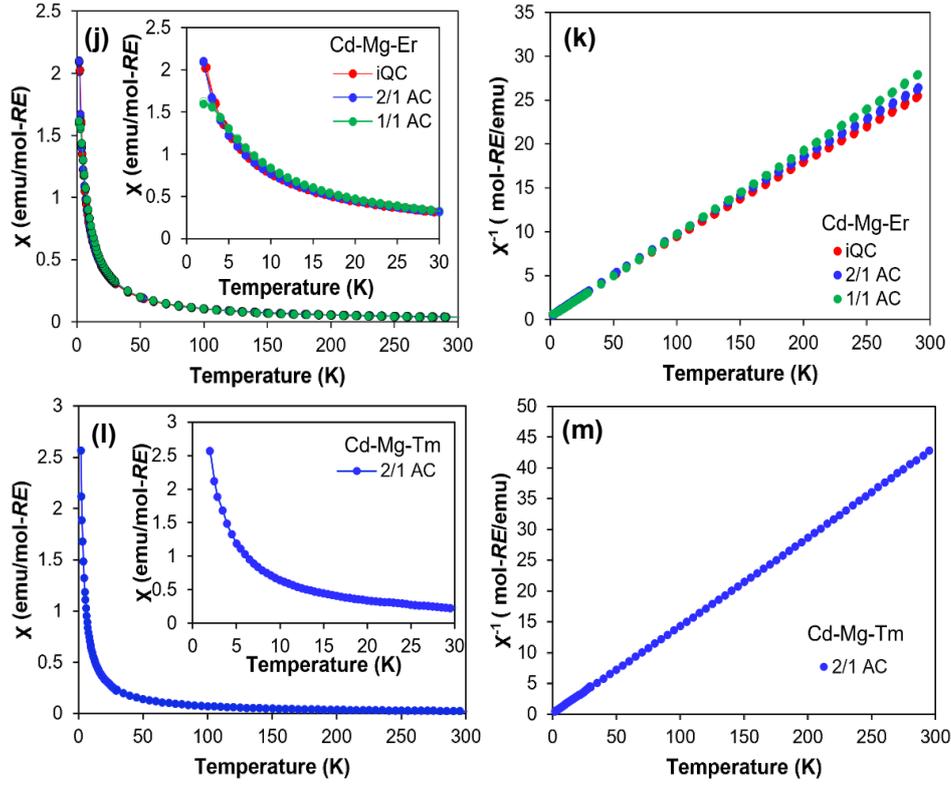

**Figure 5.** (Continued)

**Table 2.** Weiss temperature ($\Theta_w$), the freezing temperature ($T_f$), and effective magnetic moment ($\mu_{eff}$) of the i-QC, 2/1 and 1/1 ACs

| iQC/AC | Composition | $\mu_{eff}$ ($\mu_B/RE_{ion}$) | $\mu_{calc.}$ ($\mu_B/RE_{ion}$) | $\Theta_w$ (K) | $T_f$ (K) | $T_N$ (K) | $|\Theta_w/T_f|$ |
|---|---|---|---|---|---|---|---|
| i-QC | $Cd_{65}Mg_{22.5}Gd_{12.5}$ | 7.8±0.2 | 7.94 | −31.0±1.5 | 5.0±0.5 | − | 6.2±1.5 |
| 1/1 AC | $Cd_{65}Mg_{22}Gd_{13}$ | 7.6±0.2 |  | −26.5±1.5 | 12.5±0.5 | − | 2.0±1.5 |
| i-QC | $Cd_{65}Mg_{22.5}Tb_{12.5}$ | 9.8±0.2 |  | −25.5±1.5 | 6.0±0.5 | − | 5.1±1.5 |
| 2/1 AC | $Cd_{62}Mg_{25}Tb_{13}$ | 9.6±0.2 | 9.72 | −23.0±1.5 | 11.0±0.5 | − | 2.0±1.5 |
| 1/1 AC | $Cd_{65}Mg_{20}Tb_{15}$ | 9.7±0.2 |  | −17.5±1.5 | 13.0±0.5 | − | 1.3±1.5 |
| i-QC | $Cd_{65}Mg_{22.5}Dy_{12.5}$ | 10.8±0.2 |  | −17.0±2.0 | 3.0±0.5 | − | 5.5±1.5 |
| 2/1 AC | $Cd_{62}Mg_{25}Dy_{13}$ | 10.6±0.2 | 10.63 | −13.5±1.5 | 4.0±0.5 | − | 2.7±1.5 |
| 1/1 AC | $Cd_{75}Mg_{10}Dy_{15}$ | 10.0±0.2 |  | −8.0±2.0 | − | 11.0±0.5 | 0.8±1.5 |
| i-QC | $Cd_{65}Mg_{22.5}Ho_{12.5}$ | 10.7±0.2 |  | −7.0±2.0 | − | − | − |
| 2/1 AC | $Cd_{62}Mg_{25}Ho_{13}$ | 10.5±0.2 | 10.60 | −5.5±1.5 | 4.0±0.5 | − | 1.7±1.5 |
| 1/1 AC | $Cd_{75}Mg_{10}Ho_{15}$ | 10.2±0.2 |  | −4.0±1.5 | − | 7.0±0.5 | 0.8±1.5 |
| i-QC | $Cd_{65}Mg_{22.5}Er_{12.5}$ | 9.6±0.2 |  | −3.5±1.5 | − | − | − |
| 2/1 AC | $Cd_{62}Mg_{25}Er_{13}$ | 9.5±0.2 | 9.59 | −4.0±1.5 | − | − | − |
| 1/1 AC | $Cd_{75}Mg_{10}Er_{15}$ | 9.3±0.2 |  | −1.5±1.5 | − | 2.5±0.5 | 0.5±1.5 |
| 2/1 AC | $Cd_{62}Mg_{25}Tm_{13}$ | 7.2±0.2 | 7.57 | −0.5±1.5 | − | − | − |

Large $|\Theta_w/T_f|$ values have also been reported in iQCs and ACs in other alloys systems such as 10 in binary Cd-Gd [6], 4.5 in ternary Zn-Mg-Tb [35,36], 2.4-17 in ternary Ag-In-RE [37] and 2.3-3.6 in ternary Au-Al-Tm [38] systems. Larger $|\Theta_w/T_f|$ values for iQCs than ACs may be interpreted as an evidence for higher competition of the magnetic interactions in aperiodic, rather than periodic systems.

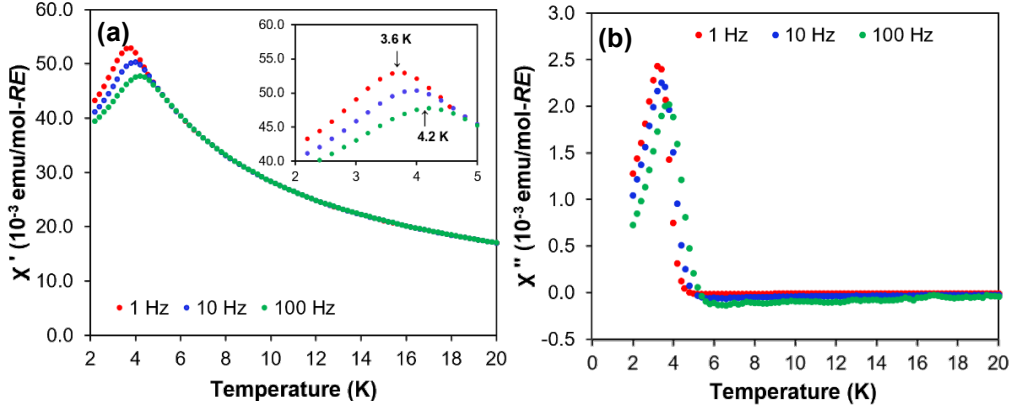

**Figure 6.** (a) The real and (b) imaginary parts of the ac susceptibility of the Cd–Mg–Ho 2/1 AC measured under $f_{ac}$ = 1 - 100 Hz.

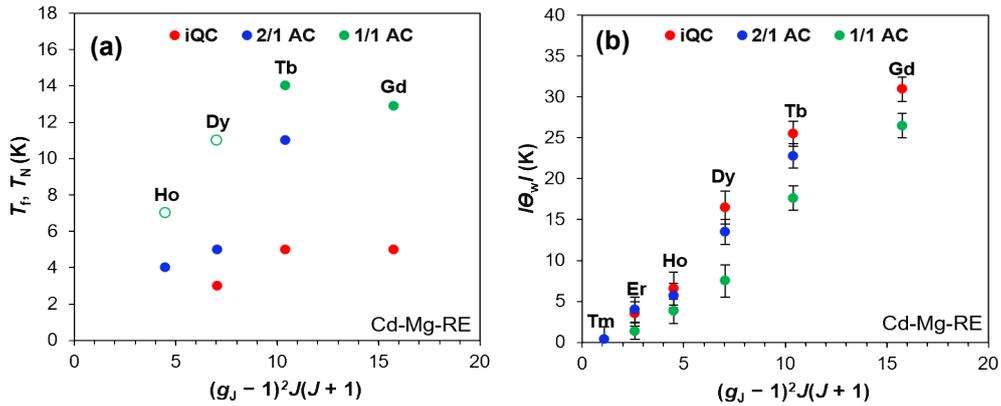

**Figure 7.** Variation of (a) $T_f$ and (b) $|\theta_w|$ as a function of the de Gennes factor: $(g_J - 1)^2 J(J + 1)$ for the magnetic iQC, 2/1 and 1/1 ACs in the Cd-Mg-*RE* systems. The open circles for 1/1 ACs in Cd-Mg-*RE* (*RE* = Ho and Dy) systems in figure 6 (a) represent the Néel temperatures. The occurrence of ordering in these two samples is associated with their lower Mg concentration of 10 at. % compared to the 1/1 ACs in Cd-Mg-*RE* (*RE* = Tb and Gd) systems that include 20-22 at. % Mg. This is due to the fact that the single domain (Cd,Mg)$_6$*RE* expands up to ~10 at. % Mg in Cd-Mg-RE (RE = Ho and Dy) systems.

To obtain some insight into the difference in the competitions of interactions between iQC and ACs, we show the multiplicity of $RE^{3+}$- $RE^{3+}$ distances and the distance dependence of Ruderman-Kittel-Kasuya-Yosida (RKKY)-type interaction in Figure 8. For indirect RKKY-type interactions, the interaction strength oscillates with distance between the spins ($r$) and fades away in sufficiently long-ranged distances as $(1/r)^3$ [23,31,39]. By assuming a free-electron model, the coupling constant $J(r)$ between spins is expressed by the following relation [39]:

$$J(r) = -9\pi \left(\frac{N}{V}\right)^2 \frac{j_0^2}{\varepsilon_F} f(2k_F r_{ij}), \quad (3)$$

where $f(x)$ is given as:

$$f(x) = (-x\cos x + \sin x)/x^4 \quad (4)$$

In equation 3, $N/V$, $j_0$, $\varepsilon_F$, $r_{ij}$ and $k_F$ denote the number of electrons per unit cell, the RKKY coupling strength, the Fermi energy, the distance between two spins and the Fermi wave vector. Under the approximation of spherical symmetry, the Fermi wave vector is determined as:

$$k_F = \left(\frac{3\pi^2 N}{V}\right)^{1/3} \quad (5)$$

which only depends on the electron concentration.

The oscillatory behavior of the RKKY interaction leads to occurrence of competing FM and AFM interactions alternatively [31]. Based on the preliminary structure analysis [40], the 1/1 and 2/1 AC unit cells contain 176 and 702 atoms, respectively. By assuming Cd and Mg to be divalent and RE elements to be trivalent (based on the obtained effective moments in Table 2), the number of free electrons within a unit cell is estimated to be 376 and 1482 in 1/1 and 2/1 ACs, respectively. Considering the lattice parameter of the 1/1 AC (resp. 2/1 AC) to be 1.54 (resp. 2.49 Å), the Fermi wave vector $k_F$ is calculated to be 1.45 x $10^{10}$ m$^{-1}$ (resp. 1.44 x $10^{10}$ m$^{-1}$). Note that the afore-mentioned structure parameters of the 1/1 and 2/1 AC belong to, respectively, Cd-Mg-Y and Cd-Mg-Er systems, as typical examples. Figure 8 plots variation of $f(2k_F r_{ij})$ and distribution of $RE^{3+}$ spins as a function of interspin distance $r_{ij}$ for 1/1 and 2/1 ACs up to 10 Å. Note that each set of the symmetrically equivalent REs in the 2/1 AC, i.e. RE1−RE5, as illustrated in Figure 2, has distinct local environment and thus exhibits different distribution of RE-RE distances in Figures 8b-f. The bin width of $\delta$ = 0.07 Å is considered for plotting the histograms of nearest-neighbor distances. While multiple $RE^{3+}$- $RE^{3+}$ distances exist in the 2/1 AC (Figure 8b-f), quite few ones occur in the 1/1 AC (Figure 8a). This larger variation of magnetic interactions in 2/1 AC may result in much pronounced competition for 2/1 AC. This explains the measured larger $|\theta_w/T_f|$ values for the 2/1 AC in Table 2 compared to those obtained for 1/1 AC. Taking into account high structural resemblance between the iQC and 2/1 AC [12], it may be reasonable to assume that the $RE^{3+}$ spin distribution and RKKY oscillation in the iQC would be similar to that of the 2/1 AC, presumably resulting in similar or even more pronounced competition of magnetic interactions in iQC. Note that similar approach using simplest free-electron model was already applied to explain composition-driven spin glass to ferromagnetic transition in the quasicrystal approximant in Au-Al-Gd system [39]. It should be mentioned that although Figure 8 captures an unevenness in the RE-RE pair distributions and thus evidences a more pronounced competition among FM and AFM exchanges in the 2/1 AC than 1/1 AC using the free-electron model, it is still deficient in depicting realistic RKKY oscillation regarding the sign of the $J(r)$. This inconsistency requires more quantitative analysis, whereby it is necessary to go beyond the simplest free-electron model.

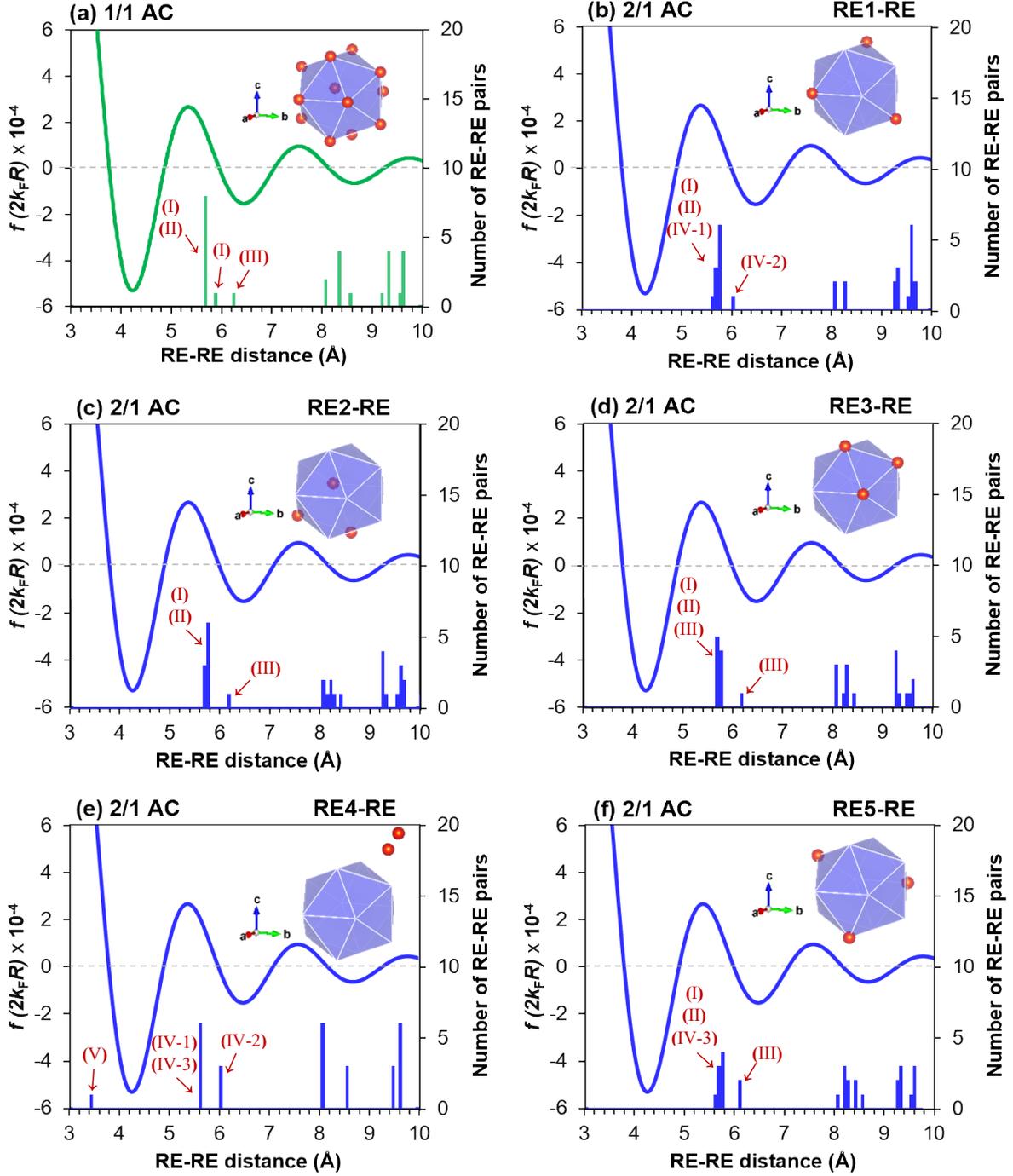

**Figure 8.** Comparison between the number of nearest-neighbour spins as a function of interspin distances for (a) 1/1 AC and (b-f) 2/1 AC. Interspin distance dependence of the RKKY interaction calculated for the 1/1 and 2/1 AC using f($x$) = ($-x \cos x + \sin x$)/$x^4$ where $x$ = $2k_F R$ is co-plotted in the same figure. $R$ stands for *RE-RE* distances. Symmetrically distinct RE sites have different distributions of RE-RE distances, thus bearing distinct local environments. The occurrences of distances in different categories are indicated within a range of 3.4 to 6.2 Å, for clarity. The results are based on the structure refinement models of the 1/1 AC (resp. 2/1 AC) in the Cd-Mg-Y (resp. Cd-Mg-Er) systems [40].

We note that to discuss the origin of the spin-glass behavior in the present samples the chemical disorder of the Cd/Mg should be taken into account. This can be inferred from the observation that the long-range antiferromagnetic order in the 1/1 ACs with less Mg composition is replaced by the spin-glass-like freezing by increasing Mg composition. One may also consider the orientational disorder of the central tetrahedron of the RTH clusters as another source of disorder contributing to the observed spin-glass-like behavior of the present samples. This kind of disorder was considered to be magnetically equivalent to the chemical disorder in iQC and AC in the ternary Au-Al-RE system [38,41], which are isostructural with those in Cd-RE system. Quite recently, the importance of the chemical disorder was also indicated directly by measuring crystalline-electric-field splitting in Au-Si-Tb 1/1 AC [42]. Therefore, it seems reasonable to assume that the combination of chemical disorder and competition between FM and AFM interactions due to oscillating RKKY polarization is responsible for the spin-glass-like feature in the present samples.

## 4. Conclusion

The present research set out to compare the magnetic properties of the iQC, 2/1 and 1/1 ACs in the ternary Cd-Mg-RE (RE = Gd−Tm) systems. The following conclusions are drawn from the present work: At higher temperatures (100 K < $T$ < 300 K), all the iQCs and ACs follow Curie–Weiss law. The estimated $\mu_{eff}$ are close to the calculated values for free $RE^{3+}$. The $\theta_w$ values are negative indicating that RE-RE exchange interactions are dominantly AFM. On the other hand, at lower temperatures, iQC and 2/1 ACs exhibit spin-glass-like anomalies for the RE atoms except Tm and Er. The 1/1 ACs exhibit either spin-glass-like freezing or AFM ordering depending on Mg content. The $T_f$ values show increasing trend from iQC to 2/1 and 1/1 ACs. In contrast, $|\theta_w|$ values for iQCs are larger than those for 2/1 and 1/1 ACs, indicating that the total AFM interactions between the neighbour spins are larger in aperiodic systems than periodic ones. The combination of chemical disorder and competition between FM and AFM interactions due to the oscillating RKKY polarization, is presumably responsible for the spin-glass-like feature that is observed in most of the present compounds.

## Acknowledgements

This work was supported in part by Japan Society for the Promotion of Science through Grants-in-Aid for Scientific Research (Grant Nos. JP19H05819, JP17K18744, JP19H01834, JP19K21839, JP19H05824 and JP19K03709). This work was also supported partly by the research program "dynamic alliance for open innovation bridging human, environment, and materials".

## References

[1] Shechtman D, Blech I, Gratias D and Cahn J W 1984 Metallic phase with long-range orientational order and no translational symmetry *Phys. Rev. Lett.* **53** 1951–3

[2] Duneau M and Katz A 1985 Quasiperiodic patterns *Phys. Rev. Lett.* **54** 2688–91

[3] Elser V and Henley C L 1985 Crystal and quasicrystal structures in Al-Mn-Si alloys *Phys. Rev. Lett.* **55** 2883–6

[4] Henley C L and Elser V 2007 Quasicrystal structure of (Al, Zn)49Mg32 *Philos. Mag. B* **53** L59–66

[5] Tsai A P, Guo J Q, Abe E, Takakura H and Sato T J 2000 A stable binary quasicrystal *Nature* **408** 537–537

[6] Goldman A I, Kong T, Kreyssig A, Jesche A, Ramazanoglu M, Dennis K W, Bud'ko S L and Canfield P C 2013 A family of binary magnetic icosahedral quasicrystals based on rare earths and cadmium *Nat. Mater.* **12** 714–8

[7] Guo J, Abe E and Tsai A P 2000 Stable Icosahedral Quasicrystals in the Cd – Mg – RE ( RE = Rare Earth Element ) Systems *Appl. Phys.* **39** 770–1

[8] Guo J Q, Abe E and Tsai A P 2002 Stable Cd-Mg-Yb and Cd-Mg-Ca icosahedral quasicrystals with wide composition ranges *Philos. Mag. Lett.* **82** 27–35

[9] Guo J Q and Tsai A P 2002 Stable icosahedral quasicrystals in the Ag–In–Ca, Ag–In–Yb, Ag–In–Ca–Mg and Ag–In–Yb–Mg systems. *Philos. Mag. Lett.* **82** 349–52

[10] Ishimasa T, Tanaka Y and Kashimoto S 2011 Icosahedral quasicrystal and 1/1 cubic approximant in Au-Al-Yb alloys *Philos. Mag.* **91** 4218–29

[11] Tanaka K, Tanaka Y, Ishimasa T, Nakayama M, Matsukawa S, Deguchi K and Sato N K 2014 Tsai-type quasicrystal and its approximant in Au-Al-Tm alloys *Acta*

*Phys. Pol. A* **126** 603–7

[12]  Takakura H, Gómez C P, Yamamoto A, De Boissieu M and Tsai A P 2006 Atomic structure of the binary icosahedral Yb–Cd quasicrystal *Nat. Mater.* **6** 58–63

[13]  Gómez C P and Lidin S 2003 Comparative structural study of the disordered MCd6 quasicrystal approximants *Phys. Rev. B* **68** 024203

[14]  Sato T J 2005 Short-range order and spin-glass-like freezing in A-Mg-R (A = Zn or Cd; R = rare-earth elements) magnetic quasicrystals *Acta Crystallogr. A* **61** 39–50

[15]  Goldman A I 2014 Magnetism in icosahedral quasicrystals: Current status and open questions *Sci. Technol. Adv. Mater.* **15** 044801

[16]  Sebastian S E, Huie T, Fisher I R, Dennis K W and Kramer M J 2004 Magnetic properties of single grain R-Mg-Cd primitive icosahedral quasicrystals (R=Y, Gd, Tb or Dy) *Philos. Mag.* **84** 1029–37

[17]  Sato T J, Takakura H, Guo J, Tsai A P, Sato T J, Takakura H, Guo J, Tsai A P and Ohoyama K 2002 Magnetic correlations in the Cd-Mg-Tb icosahedral quasicrystal *J. Alloys Compd.* **342** 365–8

[18]  Kong T, Bud S L, Jesche A, Mcarthur J, Kreyssig A, Goldman A I and Canfield P C 2014 Magnetic and transport properties of i- R -Cd icosahedral quasicrystals ( R = Y , Gd-Tm ) *Phys. Rev. B* **014424** 1–13

[19]  Yamada T, Takakura H, Kong T, Das P, Jayasekara W T, Kreyssig A, Beutier G, Canfield P C, De Boissieu M and Goldman A I 2016 Atomic structure of the i- R -Cd quasicrystals and consequences for magnetism *Phys. Rev. B* **94** 060103(R)

[20]  Fujiwara T and Ishi Y 2008 *Quasicrystals: handbook of metal physics* (Amsterdam: Elsevier)

[21]  Takakura H, Gómez C P, Yamamoto A, De Boissieu M and Tsai A P 2007 Atomic structure of the binary icosahedral Yb–Cd quasicrystal *Nat. Mater.* **6** 58–63

[22]  Tamura R, Muro Y, Hiroto T, Nishimoto K and Takabatake T 2010 Long-range magnetic order in the quasicrystalline approximant Cd6Tb *Phys. Rev. B* **82** 220201

[23]  Ori A M, Ta H O, Oshiuchi S Y, Wakawa K I, Aga Y T, Irose Y H, Akeuchi T T, Amamoto E Y and Aga Y H 2012 Electrical and Magnetic Properties of Quasicrystal Approximants RCd6 ( R: Rare Earth) *J. Phys. Soc. Jpn.* **81** 1–10

[24]  Kreyssig A, Beutier G, Hiroto T, Kim M G, Tucker G S, Boissieu M De, Tamura R and I.goldman A 2013 Antiferromagnetic order and the structural order-disorder transition in the Cd6Ho quasicrystal approximant *Philos. Mag. Lett.* **93** 512–20

[25]  Gebresenbut G, Andersson M S, Beran P, Manuel P, Nordblad P, Sahlberg M and Gomez C P 2014 Long range ordered magnetic and atomic structures of the quasicrystal approximant in the Tb-Au-Si system *J. Phys.: Condens. Matter* **26** 322202

[26]  Hiroto T, Tokiwa K and Tamura R 2014 Sign of canted ferromagnetism in the quasicrystal approximants Au-SM-R (SM = Si, Ge and Sn / R = Tb, Dy and Ho) *J. Phys.: Condens. Matter* **26** 216004

[27]  Hiroto T and Gebresenbut G H 2013 Ferromagnetism and re-entrant spin-glass transition in quasicrystal approximants Au-SM-Gd ( SM = Si , Ge ) *J. Phys.: Condens. Matter* **25** 426004

[28]  Labib F, Fujita N, Ohhashi S and Tsai A P 2020 Icosahedral quasicrystals and their cubic approximants in the Cd-Mg-RE (RE = Y, Sm, Gd, Tb, Dy, Ho, Er, Tm) systems *J. Alloys Compd.* **822** 153541

[29]  Labib F, Ohhashi S and Tsai A P 2019 Formation and crystallographic orientation study of quasicrystal, 2/1 and 1/1 approximants in Cd–Mg–Y system using electron backscatter diffraction (EBSD) *Philos. Mag.* **99** 1528–50

[30]  Tsai A P 2013 Discovery of stable icosahedral quasicrystals: progress in understanding structure and properties. *Chem. Soc. Rev.* **42** 5352–65

[31]  Mydosh J a. 1993 *Spin Glasses: An Experimental Introduction* (London: Taylor & Francis)

[32]  Wang P, Stadnik Z M, Al-Qadi K and Przewoźnik J 2009 A comparative study of the magnetic properties of the 1/1 approximant Ag50In36Gd14 and the icosahedral quasicrystal Ag50In36Gd14 *J. Phys.: Condens. Matter* **21** 436007

[33]  Labib F, Okuyama D, Fujita N, Yamada T, Ohhashi S, Sato T J, Tsuda K, Morikawa D and Tsai A P in preparation

[34]  Sato T J, Guo J and Tsai A P 2001 Magnetic properties of the icosahedral Cd-Mg-rare-earth quasicrystals *J. Phys.: Condens. Matter* **13** L105–11

[35]  Hattori Y, Niikura A, Tsai A P, Inue A, Masumoto T, Fukamichi K, Aruga-Katori H and Goto T 1995 Spin-glass behaviour of icosahedral Mg-Gd-Zn and Mg-Tb-Zn quasi-crystals *J. Phys.: Condens. Matter* **7** 2313–20

[36]  Fisher I R, Cheon K O, Panchula A F, Canfield P C, Chernikov M, Ott H R and Dennis K W 1999 Magnetic and transport properties of single-grain R -Mg-Zn icosahedral quasicrystals [R=Y, (Y1-xGdx), (Y1-xTbx), Tb, Dy, Ho, and Er] *Phys. Rev. B* **59** 308–21

[37]  Ibuka S, Iida K and Sato T J 2011 Magnetic properties of the Ag-In-rare-earth 1/1 approximants *J. Phys.: Condens. Matter* **23** 056001–8

[38]  Nakayama M, Tanaka K, Matsukawa S, Deguchi K, Imura K, Ishimasa T and Sato N K 2015 Localized electron magnetism in the icosahedral Au-Al-Tm quasicrystal and crystalline approximant *J. Phys. Soc. Jpn.* **84** 5–10

[39]  Ishikawa A, Hiroto T, Tokiwa K, Fujii T and Tamura R 2016 Composition-driven spin glass to ferromagnetic transition in the quasicrystal approximant Au-Al-Gd *Phys. Rev. B* **93** 1–6

[40]  Yamada T, Labib F and Fujita N in preparation

[41]  Matsukawa S, Tanaka K, Nakayama M, Deguchi K, Imura K, Takakura H, Kashimoto S, Ishimasa T and Sato N K 2014 Valence change driven by constituent element substitution in the mixed-valence quasicrystal and approximant Au-Al-Yb *J. Phys. Soc. Jpn.* **83** 1–5

[42]  Hiroto T, Sato T J, Cao H, Hawai T, Yokoo T, Itoh S and Tamura R 2019 Noncoplanar ferrimagnetism and local crystalline-electric-field anisotropy in the quasicrystal approximant Au70Si17Tb13 *arXiv:1912.13180* 1–14